\documentclass[prd,aps]{revtex4}

\newcommand{\nn}{\nonumber}

\begin{document}

\title{Note on ingoing coordinates for binary black holes}
\author{Kashif Alvi}
\affiliation{Theoretical Astrophysics, California Institute of Technology,
Pasadena, California 91125}

\begin{abstract}

In a previous paper, a binary black hole four-metric was presented in
a post-Newtonian corotating coordinate system valid only up to the holes'
apparent horizons.  In this paper, I
define an ingoing coordinate transformation that extends this corotating
coordinate system through the holes' horizons and into their interiors.
The motivation for using ingoing coordinates is that numerical
simulations of black holes require the computational grid to extend inside the
horizons.  The coordinate transformation presented here makes the binary
black hole four-metric suitable as a source
of initial data for numerical simulations.
\end{abstract}

\maketitle

\section{Introduction}
\label{ingintro}

In a previous paper \cite{alvi}, an approximate solution to
Einstein's equations representing two widely separated
nonrotating black holes in a circular orbit was
constructed by matching a post-Newtonian metric to two perturbed Schwarzschild
metrics.  The spacetime metric was presented in a single coordinate system
valid up to the holes' apparent horizons.  In this paper, I write the
binary black hole four-metric from \cite{alvi}
in coordinates that are corotating and post-Newtonian in the
radiation and near zones and smoothly become ingoing near the black holes.
This coordinate system is valid
through the holes' horizons and covers the holes' interiors as well as the
near and radiation zones.
The metric
components in this coordinate system are explicitly nonsingular on
the black hole horizons.  The metric presented here is promising as a source
of initial data for numerical simulations of binary black holes.
Since these
simulations require the computational grid to extend inside the
holes' horizons, the coordinate system used near the black holes in
\cite{alvi} is not suitable for numerical relativity.

Let us begin with the metric near the first black hole, that is, in region
I in the terminology of \cite{alvi} (see Fig. 1 in \cite{alvi}).
This metric is the Schwarzschild
metric plus electric-type and magnetic-type tidal
perturbations due to the second
black hole, and is given in isotropic coordinates in Eq. (3.22)
of \cite{alvi}.  The second black hole's tidal field
rotates with angular velocity $\Omega$ as seen by inertial observers in the
first black hole's local asymptotic rest frame.  However, the tidal
perturbation's angular
dependence $\phi-\Omega\tilde{t}$ as given in Eq. (3.22) of \cite{alvi}
(I have replaced $T$ in that equation with $\tilde{t}$) is
singular at the first black hole's horizon.
The reason is that the Schwarzschild time coordinate
$\tilde{t}$, which is suitable for applying the technique of
matched asymptotic expansions in the buffer zone around the black hole
(see \cite{alvi} for details),
is badly behaved at the horizon.  Since our goal in this paper
is to come up with coordinates valid through the
horizon and inside the black hole, we must use a time coordinate $T$ with
the property that the hypersurfaces of constant time coincide with
Schwarzschild time slices in the buffer zone but smoothly transition into
ingoing Eddington-Finkelstein time slices which penetrate the horizon.
The singular angular dependence $\phi-\Omega\tilde{t}$ can be simply
replaced by the nonsingular $\phi-\Omega T$, with $T$ as described above;
this is discussed in further detail below.

It turns out that isotropic coordinates are not a good starting point for
an ingoing transformation.  The analog of the ingoing Eddington-Finkelstein
transformation, which is based on ingoing null geodesics of
the Schwarzschild spacetime, is unsuccessful when applied to isotropic
coordinates: the coordinate system remains singular at the
horizon.
Indeed, the isotropic radial coordinate is only defined outside the black hole.
However, isotropic coordinates were used in \cite{alvi}
to match a tidally perturbed
Schwarzschild metric to the post-Newtonian near zone metric.
It is therefore necessary to define a new radial
coordinate that is equal to the (tidally distorted) isotropic radial
coordinate in the
buffer zone but transitions smoothly into the (tidally distorted)
Schwarzschild radial coordinate near the black hole.

\section{Ingoing coordinates}

Following the notation in \cite{alvi}, I denote the black holes'
masses by $m_1$ and $m_2$, and their coordinate separation in
post-Newtonian harmonic coordinates
by $b$.  Let $m=m_1+m_2$, $\epsilon=(m/b)^{1/2}$, and $\Omega=(1-m_1 m_2/mb)
(m/b^3)^{1/2}$.  By assumption, $\epsilon\ll 1$.

Let us begin with the region I metric given in isotropic coordinates
in Eq. (3.22) of \cite{alvi}.  Note that, in this paper,
$T$ and $R$ denote
the nonsingular time and radial coordinates described in Sec. \ref{ingintro},
while in \cite{alvi}, they denoted the isotropic time and radial
coordinates.
Set $\Omega=0$ in Eq. (3.22) of \cite{alvi} and transform
to Schwarzschild coordinates $(\tilde{t},\tilde{r},\theta,\phi)$.
This yields the metric $\tilde{\mathbf{g}}=\mathbf{g}_S+\tilde{\mathbf{h}}$;
the Schwarzschild metric $\mathbf{g}_S$ and the stationary tidal perturbation
$\tilde{\mathbf{h}}$ are given in Schwarzschild coordinates by
\begin{eqnarray}
\mathbf{g}_S &=& -\left(1-{2m_1\over \tilde{r}}\right){d\tilde{t}}^2
	+\left(1-{2m_1\over \tilde{r}}\right)^{-1}{d\tilde{r}}^2
	+{\tilde{r}}^2(d\theta^2+\sin^2\theta d\phi^2),\label{gSschw}\\
\tilde{\mathbf{h}} &=& -{4m_2\over b^3}\sqrt{{m\over b}}
	\left(1-{2m_1\over \tilde{r}}
	\right){\tilde{r}}^3 dt[\cos\theta\sin\phi d\theta
	+\sin\theta\cos(2\theta)\cos\phi d\phi]\nn\\
& &\mbox{}+{m_2{\tilde{r}}^2\over b^3}\left[3\sin^2\theta\cos^2\phi-1\right]
	\Biggl[\left(1-{2m_1\over\tilde{r}}\right)^2 dt^2+{d\tilde{r}}^2\nn\\
& &\mbox{}+({\tilde{r}}^2-2m_1^2)(d\theta^2+\sin^2\theta d\phi^2)\Biggr].
\label{tildehschw}
\end{eqnarray}
In this notation, $d\tilde{t}$, $d\tilde{r}$, $d\theta$, and $d\phi$ are
coordinate one-forms and ${d\tilde{t}}^2$ denotes the tensor product
$d\tilde{t}\otimes d\tilde{t}$.

Let $\zeta_1$ and $\zeta_2$ be two numbers satisfying $2<\zeta_1<\zeta_2
<(b/m_1)^{1/2}$.
Define the new ingoing coordinates $(T,R,\theta,\phi)$ by
\begin{eqnarray}
\tilde{t} &=& T-2m_1 \ln\left(\frac{R}{2m_1}-1\right)\psi(R),\label{Tdef}\\
\tilde{r} &=& R+m_1 \left(1+\frac{m_1}{4R}\right)\eta(R).\label{Rdef}
\end{eqnarray}
The functions $\psi(R)$ and $\eta(R)$ must satisfy the following properties:
(i) for $R\geq \zeta_2 m_1$, $\psi(R)=0$ and $\eta(R)=1$ so the
coordinates are isotropic;
(ii) for $R\leq \zeta_1 m_1$, $\psi(R)=1$ and $\eta(R)=0$ so the
time coordinate is Eddington-Finkelstein and the radial coordinate is
Schwarzschild; and
(iii) for $\zeta_1 m_1<R<\zeta_2 m_1$, $\psi(R)$ and $\eta(R)$ smoothly
and monotonically vary between their constant values outside this interval.
The transition points $R=\zeta_1 m_1$ and $R=\zeta_2 m_1$ and the functions
$\psi(R)$ and $\eta(R)$ can be chosen freely as long as the above
properties are satisfied.  Since $d\tilde{r}/dR=1+(m_1+m_1^2/4R)\eta'(R)
-m_1^2\eta(R)/4R^2\geq~1-m_1^2/4R^2$, we have $d\tilde{r}/dR>0$ for $R>2m_1$,
and so Eqs.
(\ref{Tdef}) and (\ref{Rdef}) define a valid coordinate transformation.

Transforming (\ref{gSschw}) and (\ref{tildehschw}) using (\ref{Tdef}) and
(\ref{Rdef}), we obtain
\begin{eqnarray}
\mathbf{g}_S &=& -f(dT^2-2\Psi dTdR+\Psi^2 dR^2)+{H^2\over f} dR^2
	+{\tilde{r}}^2(d\theta^2+\sin^2\theta d\phi^2),\label{gSing}\\
\tilde{\mathbf{h}} &=& -{4\epsilon m_2\over b^3}f
	{\tilde{r}}^3(dT-\Psi dR)[\cos\theta\sin\phi d\theta
	+\sin\theta\cos(2\theta)\cos\phi d\phi]\nn\\
& & \mbox{}+{m_2\over b^3}{\tilde{r}}^2(3\sin^2\theta\cos^2\phi-1)
	[f^2(dT^2-2\Psi dTdR+\Psi^2 dR^2)+H^2 dR^2\nn\\
& & \mbox{}+({\tilde{r}}^2-2m_1^2)(d\theta^2+\sin^2\theta d\phi^2)],
\label{tildehing}
\end{eqnarray}
where
\begin{eqnarray}
\Psi(R) &=& 2m_1\left[R^{-1}\left(1-{2m_1\over R}\right)^{-1}\psi(R)+
	\psi'(R)\ln\left({R\over 2m_1}-1\right)\right],\label{Psidef}\\
H(R) &=& 1+m_1\left(1+{m_1\over 4R}\right)\eta'(R)-{m_1^2\over 4R^2}\eta(R),
\label{Hdef}\\
f(R) &=& 1-{2m_1\over \tilde{r}(R)},\label{fdef}
\end{eqnarray}
and $\tilde{r}$ is given in terms of $R$ in (\ref{Rdef}).  Since the
linearized Einstein equation is a tensor equation (see, e.g., Eq. (7.5.15)
in \cite{wald}) and we have only performed
a coordinate transformation, $\tilde{\mathbf{h}}$
remains a solution to this equation
on the background $\mathbf{g}_S$.  However, the $\phi$-dependence of the
perturbation $\tilde{\mathbf{h}}$ does not correspond
to the second black hole's
rotating tidal field---we set $\Omega=0$ above.  To remedy this, we simply
change $\phi$ to $\phi-\Omega T$ in the components of $\tilde{\mathbf{h}}$,
which yields a new perturbation $\mathbf{h}$.  Note that this replacement is
not a coordinate transformation; a new tensor $\mathbf{h}$ is defined.
Also note that
$T$ becomes an ingoing coordinate near the horizon, so the time dependence
$\phi-\Omega T$ causes no problems on the horizon.
This simple remedy works for the
following reason: if we solve the linearized Einstein Eq. order by order
in $\epsilon$, then time derivatives of the components of $\mathbf{h}$
produce factors of $m_1\Omega\sim\epsilon^3$ and can thus be neglected.
The perturbation $\mathbf{h}$ is given by
\begin{eqnarray}
\mathbf{h} &=& -{4\epsilon m_2\over b^3}f
	{\tilde{r}}^3(dT-\Psi dR)[\cos\theta\sin(\phi-\Omega T)d\theta
	+\sin\theta\cos(2\theta)\cos(\phi-\Omega T)d\phi]\nn\\
& & \mbox{}+{m_2\over b^3}{\tilde{r}}^2[3\sin^2\theta\cos^2(\phi-\Omega T)-1]
	[f^2(dT^2-2\Psi dTdR+\Psi^2 dR^2)+H^2 dR^2\nn\\
& & \mbox{}+({\tilde{r}}^2-2m_1^2)(d\theta^2+\sin^2\theta d\phi^2)].
\label{hing}
\end{eqnarray}

We have now arrived at a metric $\mathbf{g}=\mathbf{g}_S+\mathbf{h}$, where
$\mathbf{g}_S$ is given in (\ref{gSing}) and $\mathbf{h}$ in (\ref{hing}),
which is valid
from the black hole's interior up into and through
the buffer zone around the hole,
and is written in coordinates that are well behaved throughout this region.

\section{Transformation to corotating coordinates}

The next step is to transform the metric $\mathbf{g}$
to corotating post-Newtonian
coordinates $(t,x,y,z)$ using the transformation given in Eqs.
(4.22) and (4.23) of \cite{alvi}.
This transformation contains a rotation
that can be performed by first defining $\varphi=\phi-\Omega T$ and
then setting $\Gamma=R\sin\theta\cos\varphi$, $\Lambda=R\sin\theta\sin\varphi$,
and $Z=R\cos\theta$.  To complete the
transformation, define the functions $P_{\alpha\beta}(x,y,z)$ for
$\alpha,\beta=0,..,3$ to be components of the metric $\mathbf{g}$ in
coordinates $(T,\Gamma,\Lambda,Z)$; write these components
as functions of $(x,y,z)$ using Eqs. (4.22) and (4.23) in \cite{alvi}.
The functions $P_{\alpha\beta}$ are given by
\begin{eqnarray}
P_{00} &=& -f+\frac{{\tilde{r}}^2}{R^2}\Omega^2(\Gamma^2+\Lambda^2)
	-{4\epsilon m_2 {\tilde{r}}^3\over b^3 R^3}(2Z^2-R^2)f\Omega\Gamma\nn\\
	& &\mbox{}+{m_2 {\tilde{r}}^2\over b^3 R^2}(3\Gamma^2-R^2)\left[f^2+
	\frac{\Omega^2}{R^2}({\tilde{r}}^2-2m_1^2)({\Gamma}^2+{\Lambda}^2)
	\right],\nn\\
P_{01} &=& P_{10} = \frac{{\Gamma}}{R}f{\Psi}-\frac{{\tilde{r}}^2}{R^2}
	\Omega\Lambda
	-{2\epsilon m_2 {\tilde{r}}^3\over b^3 R^3}f\left[{\Gamma}{\Lambda}
	-\frac{\Omega}{R}{\Psi}{\Gamma}^2
	(2Z^2-R^2)\right]\nn\\
	& &\mbox{}+{m_2 {\tilde{r}}^2\over b^3 R^2}(3\Gamma^2-R^2)\left[-\frac{{\Gamma}}{R}
	f^2 {\Psi}-\frac{\Omega {\Lambda}}{R^2}({\tilde{r}}^2-2m_1^2)\right],\nn\\
P_{02} &=& P_{20} = \frac{{\Lambda}}{R}f{\Psi}+\frac{{\tilde{r}}^2}{R^2}\Omega {\Gamma}
	-{2\epsilon m_2 {\tilde{r}}^3\over b^3 R^3}f
	\left[Z^2-{\Gamma}^2-\frac{\Omega}{R}{\Psi}{\Gamma}{\Lambda}
	(2Z^2-R^2)\right]\nn\\
	& &\mbox{}+{m_2 {\tilde{r}}^2\over b^3 R^2}(3\Gamma^2-R^2)\left[-\frac{{\Lambda}}{R}
	f^2 {\Psi}+\frac{\Omega {\Gamma}}{R^2}({\tilde{r}}^2-2m_1^2)\right],\nn\\
P_{03} &=& P_{30} = \frac{Z}{R}f{\Psi}
	-{2\epsilon m_2{\tilde{r}}^3\over b^3 R^3}f\left[-{\Lambda}Z-\frac{\Omega}{R}{\Psi}{\Gamma}Z
	(2Z^2-R^2)\right]\nn\\
	& &\mbox{}+{m_2 {\tilde{r}}^2\over b^3 R^2}(3\Gamma^2-R^2)\left(-\frac{Z}{R}
	f^2 {\Psi}\right),\nn\\
P_{11} &=& \frac{{\Gamma}^2}{R^2}\left(-f{\Psi}^2+{H^2\over f}-\frac{{\tilde{r}}^2}{R^2}\right)
	+\frac{{\tilde{r}}^2}{R^2}
	+{4\epsilon m_2 {\tilde{r}}^3\over b^3 R^4}f{\Psi}{\Gamma}^2 {\Lambda}\nn\\
	& &\mbox{}+{m_2 {\tilde{r}}^2\over b^3 R^2}(3\Gamma^2-R^2)\left[
	\frac{{\Gamma}^2}{R^2}(f^2 {\Psi}^2+H^2)+
	R^{-2}({\tilde{r}}^2-2m_1^2)\left(1-\frac{{\Gamma}^2}{R^2}\right)\right],\nn\\
P_{22} &=& \frac{{\Lambda}^2}{R^2}\left(-f{\Psi}^2+{H^2\over f}-\frac{{\tilde{r}}^2}{R^2}\right)
	+\frac{{\tilde{r}}^2}{R^2}
	+{4\epsilon m_2 {\tilde{r}}^3\over b^3 R^4}f{\Psi}{\Lambda}(Z^2-{\Gamma}^2)\nn\\
	& &\mbox{}+{m_2 {\tilde{r}}^2\over b^3 R^2}(3\Gamma^2-R^2)\left[
	\frac{{\Lambda}^2}{R^2}(f^2 {\Psi}^2+H^2)+
	R^{-2}({\tilde{r}}^2-2m_1^2)\left(1-\frac{{\Lambda}^2}{R^2}\right)\right],\nn\\
P_{33} &=& \frac{Z^2}{R^2}\left(-f{\Psi}^2+{H^2\over f}-\frac{{\tilde{r}}^2}{R^2}\right)
	+\frac{{\tilde{r}}^2}{R^2}
	-{4\epsilon m_2 {\tilde{r}}^3\over b^3 R^4}f{\Psi}Z^2 {\Lambda}\nn\\
	& &\mbox{}+{m_2 {\tilde{r}}^2\over b^3 R^2}(3\Gamma^2-R^2)\left[
	\frac{Z^2}{R^2}(f^2 {\Psi}^2+H^2)+
	R^{-2}({\tilde{r}}^2-2m_1^2)\left(1-\frac{Z^2}{R^2}\right)\right],\nn\\
P_{12} &=& P_{21} = \frac{{\Gamma}{\Lambda}}{R^2}\left(-f{\Psi}^2+{H^2\over f}
	-\frac{{\tilde{r}}^2}{R^2}\right)
	+{2\epsilon m_2 {\tilde{r}}^3\over b^3 R^4}f{\Psi}{\Gamma}
	(-{\Gamma}^2+{\Lambda}^2+Z^2)\nn\\
	& &\mbox{}+{m_2 {\tilde{r}}^2\over b^3 R^4}(3\Gamma^2-R^2)\left[
	f^2 {\Psi}^2+H^2-R^{-2}({\tilde{r}}^2-2m_1^2)\right]
	{\Gamma}{\Lambda},\nn\\
P_{13} &=& P_{31} = \frac{{\Gamma}Z}{R^2}\left(-f{\Psi}^2+{H^2\over f}
	-\frac{{\tilde{r}}^2}{R^2}\right)\nn\\
	& &\mbox{}+{m_2 {\tilde{r}}^2\over b^3 R^4}(3\Gamma^2-R^2)\left[
	f^2{\Psi}^2+H^2-R^{-2}({\tilde{r}}^2-2m_1^2)\right]\Gamma Z,\nn\\
P_{23} &=& P_{32} = \frac{{\Lambda}Z}{R^2}\left(-f{\Psi}^2+{H^2\over f}
	-\frac{{\tilde{r}}^2}{R^2}\right)
	+{2\epsilon m_2 {\tilde{r}}^3\over b^3 R^4}f{\Psi}Z
	(-{\Gamma}^2-{\Lambda}^2+Z^2)\nn\\
	& &\mbox{}+{m_2 {\tilde{r}}^2\over b^3 R^4}(3\Gamma^2-R^2)\left[
	f^2 {\Psi}^2+H^2-R^{-2}({\tilde{r}}^2-2m_1^2)\right]{\Lambda}Z,\nn\\
\label{ingP}
\end{eqnarray}
where $\epsilon=(m/b)^{1/2}$, $\Omega=(1-m_1 m_2/mb)(m/b^3)^{1/2}$,
$R=(\Gamma^2+\Lambda^2+Z^2)^{1/2}$, and $\Gamma$, $\Lambda$,
and $Z$ are to be expressed in terms of $(x,y,z)$ via Eqs. (4.22)
and (4.23) in \cite{alvi}.  The functions $\tilde{r}(R)$, $\Psi(R)$,
$H(R)$, and $f(R)$ are given in Eqs. (\ref{Rdef}), (\ref{Psidef}),
(\ref{Hdef}), and (\ref{fdef}).

The remainder of the coordinate transformation from black hole coordinates
$(T,R,\theta,\phi)$ to corotating post-Newtonian coordinates $(t,x,y,z)$
can be done exactly as in \cite{alvi}.  The final metric in region I
(see Fig. 1 in \cite{alvi}) is
given by Eq. (4.27) of \cite{alvi},
but with $P_{\alpha\beta}$ taken
from Eq. (\ref{ingP}) above.
Define $\bar{P}_{\alpha\beta}$ to be $P_{\alpha\beta}$
(as given in (\ref{ingP}))
with $m_1$ and $m_2$ exchanged.  Then the
metric in region II (that is, near the second black hole; see Fig. 1 in
\cite{alvi}) is given by
Eq. (4.28) in \cite{alvi}, but with $\bar{P}_{\alpha\beta}$ taken
from here.  Note that the final metric components are everywhere explicitly
independent of time $t$.

To summarize, \textit{the expressions for the metric components given in Sec. V
of \cite{alvi} are
valid in ingoing coordinates if $P_{\alpha\beta}$ (and $\bar{P}_{\alpha\beta}$)
are taken from Eq. (\ref{ingP}) and the functions $\tilde{r}(R)$, $\Psi(R)$,
$H(R)$, and $f(R)$ are taken from Eqs. (\ref{Rdef}), (\ref{Psidef}),
(\ref{Hdef}), and (\ref{fdef}) with $\psi(R)$ and $\eta(R)$ having
properties (i)--(iii) given below Eq. (\ref{Rdef})}.
In order to show explicitly that
these components are nonsingular at the black holes' horizons, I write out
$P_{\alpha\beta}$ for $R\leq\zeta_1 m_1$.  In this region, $\psi=1$,
$\Psi(R)=2m_1 R^{-1}(1-2m_1/R)^{-1}$, $\eta=0$, $H=1$, $\tilde{r}=R$,
and $f(R)=1-2m_1/R$.  Therefore, for $R\leq\zeta_1 m_1$, $P_{\alpha\beta}$
are given by
\begin{eqnarray}
P_{00} &=& -1+{2m_1\over R}+\Omega^2({\Gamma}^2+{\Lambda}^2)
	-{4\epsilon m_2\over b^3}\left(1-{2m_1\over R}\right)
	(2Z^2-R^2)\Omega\Gamma\nn\\
& &\mbox{}+{m_2\over b^3}(3{\Gamma}^2-R^2)\left[\left(1-{2m_1\over R}\right)^2
	+\Omega^2\left(1-{2m_1^2\over R^2}\right)
	({\Gamma}^2+{\Lambda}^2)\right],\nn\\
P_{01} &=& P_{10} = {2m_1\over R^2}{\Gamma}-\Omega {\Lambda}
	-{2\epsilon m_2\over b^3}\left[\left(1-{2m_1\over R}\right)
	{\Gamma}{\Lambda}
	-{2m_1\over R^2}\Omega {\Gamma}^2(2Z^2-R^2)\right]\nn\\
& &\mbox{}-{m_2\over b^3}(3{\Gamma}^2-R^2)\left[{2m_1\over R^2}
	\left(1-{2m_1\over R}\right){\Gamma}
	+\Omega {\Lambda}\left(1-{2m_1^2\over R^2}\right)\right],\nn\\
P_{02} &=& P_{20} = {2m_1\over R^2}{\Lambda}+\Omega {\Gamma}
	-{2\epsilon m_2\over b^3}\left[\left(1-{2m_1\over R}\right)
	(Z^2-{\Gamma}^2)
	-{2m_1\over R^2}\Omega {\Gamma}{\Lambda}(2Z^2-R^2)\right]\nn\\
& &\mbox{}-{m_2\over b^3}(3{\Gamma}^2-R^2)\left[{2m_1\over R^2}
	\left(1-{2m_1\over R}\right){\Lambda}
	-\Omega {\Gamma}\left(1-{2m_1^2\over R^2}\right)\right],\nn\\
P_{03} &=& P_{30} = {2m_1\over R^2}Z
	+{2\epsilon m_2\over b^3}\left[\left(1-{2m_1\over R}\right){\Lambda}Z
	+{2m_1\over R^2}\Omega {\Gamma}Z(2Z^2-R^2)\right]\nn\\
& &\mbox{}-{2m_1 m_2\over b^3 R^2}(3{\Gamma}^2-R^2)
	\left(1-{2m_1\over R}\right)Z,\nn\\
P_{11} &=& 1+{2m_1 {\Gamma}^2\over R^3}
	+{8\epsilon m_1 m_2\over b^3 R^2}{\Gamma}^2 {\Lambda}
	+{m_2\over b^3}(3{\Gamma}^2-R^2)
	\left[1-{2m_1^2\over R^2}\left(1-{3{\Gamma}^2\over
	R^2}\right)\right],\nn\\
P_{22} &=& 1+{2m_1 {\Lambda}^2\over R^3}+{8\epsilon m_1 m_2\over b^3 R^2}
	{\Lambda}(Z^2-{\Gamma}^2)
	+{m_2\over b^3}(3{\Gamma}^2-R^2)\left[1-{2m_1^2\over R^2}
	\left(1-{3{\Lambda}^2\over
	R^2}\right)\right],\nn\\
P_{33} &=& 1+{2m_1 Z^2\over R^3}-{8\epsilon m_1 m_2\over b^3 R^2}Z^2 {\Lambda}
	+{m_2\over b^3}(3{\Gamma}^2-R^2)\left[1-{2m_1^2\over R^2}
	\left(1-{3Z^2\over R^2}\right)\right],\nn\\
P_{12} &=& P_{21} = {2m_1 {\Gamma}{\Lambda}\over R^3}
	+{4\epsilon m_1 m_2\over b^3 R^2}{\Gamma}
	(R^2-2{\Gamma}^2)+{6m_1^2 m_2\over b^3 R^4}(3{\Gamma}^2-R^2)
	{\Gamma}{\Lambda},\nn\\
P_{13} &=& P_{31} = {2m_1 {\Gamma}Z\over R^3}
	+{6m_1^2 m_2\over b^3 R^4}(3{\Gamma}^2-R^2){\Gamma}Z,\nn\\
P_{23} &=& P_{32} = {2m_1 {\Lambda}Z\over R^3}
	+{4\epsilon m_1 m_2\over b^3 R^2}Z
	(2Z^2-R^2)+{6m_1^2 m_2\over b^3 R^4}(3{\Gamma}^2-R^2){\Lambda}Z,
\label{ingPnearBH}
\end{eqnarray}
where $\epsilon=(m/b)^{1/2}$, $\Omega=(1-m_1 m_2/mb)(m/b^3)^{1/2}$,
$R=(\Gamma^2+\Lambda^2+Z^2)^{1/2}$, and $\Gamma$, $\Lambda$,
and $Z$ are to be expressed in terms of $(x,y,z)$ via Eqs. (4.22)
and (4.23) in \cite{alvi}.  Note that the quantities
$P_{\alpha\beta}$ in Eq. (\ref{ingPnearBH})
are all finite at the horizon $R=2m_1$.

\section*{Acknowledgments}

I thank Lee Lindblom, Mark Scheel, Kip Thorne, and Michele Vallisneri
for useful discussions.
This research was supported in part by NSF grant PHY-9900776.

\end{document}